\documentclass[prl,showpacs,preprintnumbers,preprint]{revtex4}

\usepackage{graphicx}

\begin{document}

\title{Zero-lag long-range synchronization via dynamical relaying}

\author
{Ingo Fischer$^{1}$, Ra\'{u}l Vicente$^{2}$, Javier M. Buld\'{u}$^{3}$, Michael Peil$^{4}$, Claudio R. Mirasso$^{2}$,\\ M. C. Torrent$^{3}$ and Jordi Garc\'{\i}a-Ojalvo$^{3}$}

\affiliation{$^{1}$Department of Applied Physics and Photonics, Vrije Universiteit Brussel, Pleinlaan 2, B-1050 Brussel, Belgium\\
$^{2}$Departament de F\'{\i}sica, Universitat de les Illes Balears, E-07071 Palma de Mallorca, Spain\\
$^{3}$Departament de F\'{\i}sica i Enginyeria Nuclear, Universitat Polit\`ecnica de Catalunya,Colom 11, 08222 Terrassa, Spain\\
$^{4}$Institute of Applied Physics, Darmstadt University of Technology, Schlossgartenstrasse 7, D-64289 Darmstadt, Germany}

\date{\today}

\begin{abstract}
We show that simultaneous synchronization between two delay-coupled oscillators can be achieved by relaying the dynamics via a third mediating element, which surprisingly lags behind the synchronized outer elements. The zero-lag synchronization thus obtained is robust over a considerable parameter range. We substantiate our claims with experimental and numerical evidence of these synchronization solutions in a chain of three coupled semiconductor lasers with long inter-element coupling delays. The generality of the mechanism is validated in a neuronal model with the same coupling architecture. Thus, our results show that synchronized dynamical states can occur over long distances through relaying, without restriction by the amount of delay.
\end{abstract}

\maketitle

Mutual coupling of oscillators often gives rise to collective dynamical behavior. Synchronization is a fascinating example of such emerging dynamics \cite{boccaletti02,pikovsky03} that plays important functional roles in complex systems. In the brain, for instance, synchronization of neural activity has been shown to underlie cognitive acts \cite{rodriguez97}. Interestingly, isochronal synchronization (without lag) can occur in the brain between widely separated cortical regions \cite{Singer1991,Singer1997}.  Near-zero delay synchronization between pairwise recordings of neuronal signals has also been recently reported \cite{Nikolic2006}. The mechanism of this phenomena has been subject of controversial debate, also in a more general context, for many years: how can two distant dynamical elements synchronize at zero lag even in the presence of non-negligible delays in the transfer of information between them? In the neuronal case, complex mechanisms and neural architectures have been proposed to answer this question \cite{traub96,koenig95,bibbig02}. However, they exhibit limitations in the maximum synchronization range (see e.g. \cite{koenig95}), and rely on complex network architectures \cite{bibbig02}.

Here we present a configuration that can be regarded as a simple network module with delayed interactions that exhibits zero-lag synchronization between two outer elements in a natural way. The module consists of three similar dynamical elements coupled bidirectionally along a line, in such a way that the central element acts as a {\em relay} of the dynamics between the outer elements. This type of network module can be expected to exist, for instance, within the complex functional architecture of the brain \cite{eguiluz:018102}.

\begin{figure}[!htb]
\centerline{
\includegraphics[width=8cm]{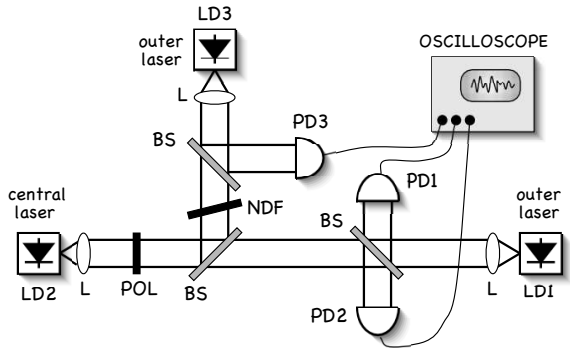}
} \caption[sdh]{Experimental setup. A central laser (LD2) exchanges information between the other two (LD1 and LD3). The coupling times between the central and outer lasers are matched between both branches.} \label{fig:setup}
\end{figure}

We have chosen semiconductor lasers for our study, since they have proven to be excellent model systems to investigate the behavior of delay-coupled elements. An advantage of semiconductor laser experiments is that these lasers can be well controlled, and that their dynamical behavior can be accurately described by established models. In addition, delays in the coupling occur generically, due to their fast dynamical timescales. The experimental setup is depicted in Fig.~\ref{fig:setup}. A central diode laser (LD2) is bidirectionally coupled to two outer lasers (LD1) and (LD3) by mutual injection. The central laser, which does not need to be carefully matched to the other two, mediates their dynamics.
The lasers (Roithner RLT6505G) have a nominal wavelength of $655$~nm, and have been coupled via their lasing TE-polarized fields. These off-the-shelf lasers exhibit typical parameter deviations of the order of a few percent, which highlights the robustness of the synchronization mechanism which we report in the following.
In order to avoid influence from the non-lasing TM modes, we have placed a polarizer (POL) before the input of laser 2. An aspheric lens (L) collimates the output beam of each laser. $50\%$ beamplitters (BS) allow to send fractions of the light to the other lasers and the rest to the corresponding photodetectors (PD). The coupling strength, i.e. the amount of light interchanged by the lasers, has been controlled via a neutral density filter (NDF). The lasers have been pumped slightly above their lasing thresholds ($I_{dc}\sim 1.07 I_{th}$) and the pump current and temperature have been controlled with high precision controllers ($\Delta I<0.01$~mA and $\Delta T<0.01$~K).

A similar coupling configuration was proposed in Ref.~\cite{winful:1575}, studying three laterally coupled lasers without delay. Experimental work in such a configuration \cite{terry:4036} confirmed that identical synchronization between the outer lasers, but not with the central one, is possible. In our case, however, the lasers are physically separated, with the outer lasers placed at similar distances of $\sim1.1$~m from the central one. This results in equal coupling times of $\tau_c \sim 3.65$~ns, corresponding to the time the light takes to propagate from one laser to the other. This time is longer than the characteristic relaxation oscillation period of the carrier-photon system of each laser, which lies in the range of 1~ns or below. Due to this delay the system becomes fundamentally different to those discussed in Refs. \cite{winful:1575,terry:4036}; delay renders the system infinite dimensional, and determines the dynamical behavior, as well as the synchronization properties and timings.

Without coupling, the three lasers emit constant power. Due to the mutual injection, the lasing threshold current of the lasers is reduced by $5-10\%$. We note that small amounts of optical feedback due to reflections at the external facet of the respective opposite laser cannot be avoided, nevertheless we have experimentally verified that they do not play an essential role. The laser outputs are sent to an optical spectrum analyzer with a resolution of $0.05$ nm, and detected by fast photodetectors ($12$~GHz bandwidth) whose signal is recorded and analyzed by a $4$~GHz oscilloscope.

If we block the beam between the central laser and one of the outer lasers, the system reduces to the case of two mutually injected lasers. This situation has been extensively studied: For short coupling delays the coupled system exhibits multistable locking for small detuning and self-sustained oscillations for large detuning \cite{PRL05}. For long delays the stable locking is lost, and coupling induces dynamical instabilities. The resulting dynamics can be synchronized between the two lasers, although in a generalized way: the lasers show similar but non-identical behavior. In particular, they are delayed with respect to each other by the coupling time. Under detuned operation, the laser with higher optical frequency leads the dynamics, while for zero detuning the two lasers spontaneously switch leader and laggard roles \cite{hei01}. The isochronous and identical synchronized solution exists mathematically, but has been found to be unstable \cite{mul04}.

\begin{figure}[!htb]
\centerline{
\includegraphics[width=7cm]{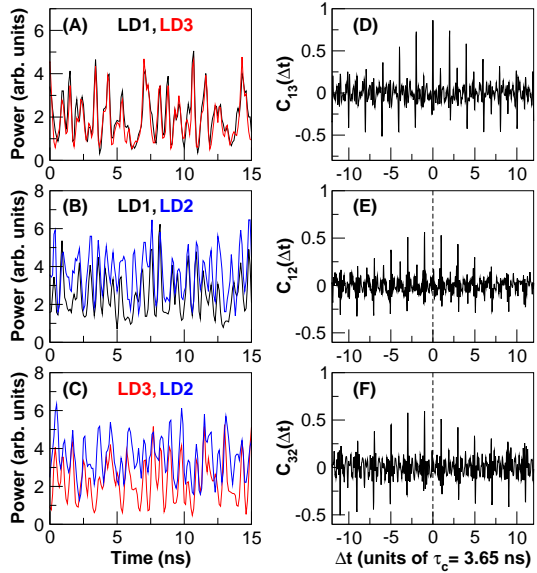}
} \caption[sdh]{\small (A-C) Time series (in pairs) of the output intensity of the lasers, for the case of a central laser with negative detuning $\Omega=\omega_{2}-\omega_{1,3}=-4.1$~GHz. (D-F) Cross-correlation functions of the corresponding time series. The time series of the central laser have been shifted $\tau_c$ to allow an easier comparison.} \label{fig:fig02}
\end{figure}

When the blocking of the isolated laser is removed, all three mutually coupled lasers exhibit chaotic outputs. Remarkably, now both outer lasers synchronize with zero lag, while the central laser either leads or lags the outer lasers. Figure~\ref{fig:fig02} shows the time series of the output intensities (left column), in pairs, and the corresponding cross-correlation functions $C_{ij}(\Delta t)$, defined as in \cite{mul04}, in such a way that a maximal cross-correlation at a positive time difference $\Delta t_{\rm max}$ indicates that element $j$ is {\em leading} element $i$ by the time $\Delta t_{\rm max}$, and vice versa.
For optimal synchronization quality, the optical frequency of the central laser has been slightly decreased with respect to the outer lasers (negatively detuned) by adjusting its temperature. Zero-lag synchronization between the intensities of the outer lasers can be clearly seen in Fig.~\ref{fig:fig02}(A), and also manifests itself in the cross-correlation function shown in Fig.~\ref{fig:fig02}(D), which presents an absolute maximum of $0.86$ at $\Delta t_{\rm max}=0$ (i.e. at zero delay). The correlation between the central laser and the outer ones [Fig.~\ref{fig:fig02}(B,C)] is not as high, and exhibits a non-zero time lag, as can be seen from the cross-correlation functions shown in Fig.~\ref{fig:fig02}(E)-(F), which yield maxima of $0.56$ and $0.59$, respectively, placed at $\Delta t_{\rm max}=-3.65$~ns. This lag coincides with the coupling time $\tau_c$ between the lasers. The fact that $\Delta t_{\rm max}$ is negative means that the central laser dynamically lags the two outer lasers. Therefore, it can be excluded that the outer lasers are simply driven by the central one. It is worth mentioning that the zero-lag synchronization is quite robust against spectral detuning of the lasers. We note that the synchronization also remains robust for positive detuning. However, then the central laser leads the dynamics.

\begin{figure}[!htb]
\centerline{
\includegraphics[width=7cm]{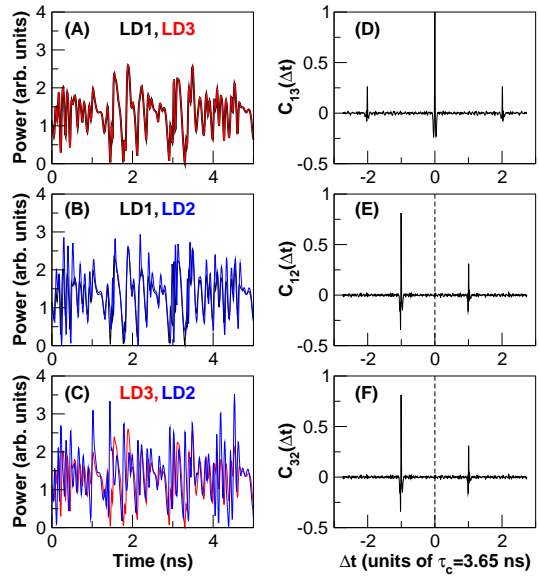}
} \caption[sdh]{\small (A-C) Numerical time series (in pairs) of the output intensity of the lasers, for the case of zero detuning between the three lasers. (D-F) Cross-correlation functions of the corresponding time series. The time series of the central laser have been shifted $\tau_c$ to allow an easier comparison.}  \label{fig:sim}
\end{figure}

In order to gain further insight into the zero-lag synchronization phenomenon and its stability, we have mathematically modeled the laser system via a rate equation model that describes the complex amplitudes of the optical fields and the corresponding carrier numbers of the three lasers:
\begin{eqnarray}
\dot{E}_m(t) & = & \frac{1}{2} (1+i \alpha) \left[ G_m - \gamma \right]
E_m(t)\nonumber\\
& + & \sum_{l=1}^{3} \kappa_{l,m} e^{-i \omega_0 \tau} E_l(t-\tau)\\
\dot{N}_m(t) & = & \frac{I_m}{e} - \gamma_e N_m - G_m |E_m|^2
\end{eqnarray}
with $G_m = g\cdot(N_m - N_0)/(1+\epsilon |E_m|^2)$.
The internal laser parameters are assumed identical for the three lasers, with linewidth enhancement factor $\alpha=3$, differential gain $g=1.2\times10^{-5}$ ns$^{-1}$, transparency inversion $N_{0}=1.25\times10^{8}$, saturation coefficient $\epsilon=5\times10^{-7}$, photon decay rate $\gamma=496$ ns$^{-1}$, carrier decay rate $\gamma_{e}=0.651$ ns$^{-1}$, $\omega_0$  being the free-running frequency of the lasers and $e$ the elementary charge. The coupling strengths, delay times and phases are assumed to be identical for the two branches of the network module: $\kappa_{1,2}=\kappa_{2,1}=\kappa_{2,3}=\kappa_{3,2}=20$~ns$^{-1}$ with $\kappa_{l,m} = 0$ otherwise, and $\tau_{1,2}=\tau_{2,1}=\tau_{2,3}=\tau_{3,2}=3.65$~ns.
The model is an extension of the one introduced in \cite{hei01} and justified in \cite{MuletPRA:02} for the case of two lasers. Here, we first discuss the perfectly symmetric situation where the lasers are identical with respect to internal parameters and operating conditions. This allows to verify whether the role of the central laser depends on asymmetries in the system. Additionally, we consider high pump currents for the lasers, a dynamical situation difficult to analyze experimentally due to the broad bandwidth of the dynamics. Figure~\ref{fig:sim}(A-C) shows time series of the two outer lasers (LD1 and LD3) and the relay laser (LD2) in pairs. One can easily notice that the dynamics of the outer lasers are more similar to each other than to that of the central one. To better analyze the dynamics we compute the cross-correlation functions between laser pairs. The results [see Fig.~\ref{fig:sim}(D-F)] show that the maximum correlation occurs at different times for different pairs.
Even in this case of zero detuning between the three lasers, the relay laser clearly lags the dynamics with respect to the outer lasers, that are almost perfectly synchronized at zero lag. As in the experiments, and even more emphasized in this case of higher injection current, the correlation is practically 1 between the outer lasers and lower between each outer laser and the central one. We note that numerical results for lower injection currents, matching the experimental results shown above, exhibit similar zero-lag synchronization properties.
In addition we have systematically studied, by means of numerical simulations, the influence of parameter mismatch of the lasers, mismatch of the delay times, and influence of external perturbations. We find that the central laser can have large parameter mismatch without preventing the occurrence of zero-lag synchronization. For relative mismatch between the outer lasers synchronization is also robust, although the acceptable tolerances are smaller in this case. Applying external perturbations via the pump current did not destroy the synchronization. Even more, we have verified that zero-lag synchronization also maintains for pump current modulation of the lasers and, which is relevant for real networks, synchronization with almost zero lag persists even for small mismatch of the coupling delay times between the lasers.

In order to prove whether the behavior is indeed generic, we have performed simulations of three Hodgkin-Huxley-type neurons connected according to the same network architecture. We have chosen a model of a thermoreceptor neuron \cite{braun} that exhibits a variety of dynamical behavior ranging from regular spiking to bursting and self-sustained chaotic pulsations, depending on the temperature. The neurons are mutually coupled in pairs via synaptic connections. The delay in the information transmission between the neurons is taken to be much longer than the internal time scale of the spiking process.

\begin{figure}[!htb]
\centerline{
\includegraphics[width=7cm]{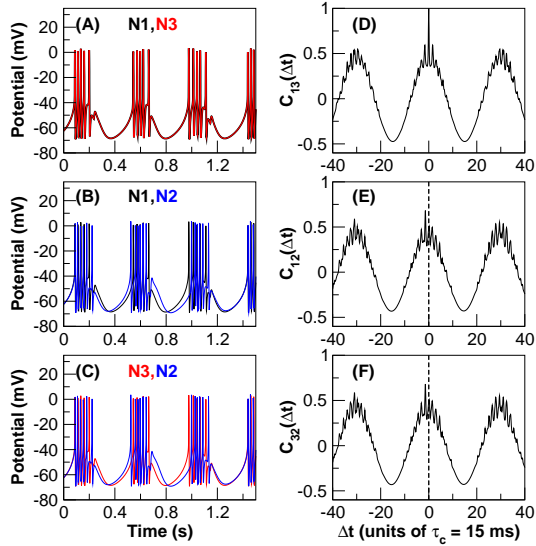}
} \caption[sdh]{\small Synchronization of three bidirectionally coupled thermoreceptor neurons. The left column shows the time series of the three neurons, and the right column the corresponding cross-correlation functions. The maxima of the cross correlation for panels E and F is about 20 ms which roughly corresponds to the coupling time. Following the notation of Ref. \protect\cite{braun}, the parameters are: $g_{\rm Na}=1.5\,\mu$S/cm$^2$,
  $g_{\rm K}=2\,\mu$S/cm$^2$,
 $g_{\rm sd}=0.25\,\mu$S/cm$^2$,
 $g_{\rm sr}=0.4\,\mu$S/cm$^2$,
  $g_{\rm l}=0.1\,\mu$S/cm$^2$,
  $g_{\rm syn}=0.15\,\mu$S/cm$^2$,
 $V_{\rm Na}=50$~mV,
  $V_{\rm K}=-90$~mV,
 $V_{\rm sd}=50$~mV,
 $V_{\rm sr}=-90$~mV,
 $V_{\rm l}=-60$~mV,
 $V_{\rm syn}=0$~mV,
 $\tau_{\rm Na}=0\,$ms,
 $\tau_{\rm K}=2.0\,$ms,
 $\tau_{\rm sd}=10.0\,$ms,
 $\tau_{\rm sr}=20.0\,$ms,
 $\tau_{\rm syn}=5\,$ms,
 $\tau_{\rm c}=15\,$ms,
 $g_i^{syn}=-0.13\,$ms$^{-1}$,
  $C_m=1\,\mu$F/cm$^2$,
 $\alpha=0.012\, \mu$A, and
 $\beta=0.17$.
} \label{fig:3hh}
\end{figure}

In the following, we have considered a regime in which isolated neurons exhibit irregular spikes grouped in regular bursts. For a network of two delay-coupled neurons, small correlations between spikes are observed at time differences corresponding to the connection time. When a mediating neuron is added, identical synchronization at zero-lag appears between the outer neurons, as shown in Fig.~\ref{fig:3hh}. Moreover, the correlation between the mediating neuron and the outer ones is significantly smaller, with the central neuron lagging the dynamics at a time that amounts to the connection delay. Therefore, the central results obtained for coupled lasers also stand for the neuron model: for two elements leader-laggard dynamics is observed, while for three elements zero-lag synchronization of the outer elements and lagging of the central element occur.

We note, that we have been able to find the presented zero-lag synchronization mechanism additionally in model calculations for a large variety of dynamical systems including excitable systems, oscillators and maps for periodic or even chaotic dynamics, proving its generic nature. In addition, it proves that the topology governs the described synchronization properties. Certain commonly occurring network modules, called motifs, have been proposed as basic building blocks of complex networks \cite{Milo02}. Those studies considered instantaneous coupling, but the fact that interactions propagate at finite speed can not always be neglected, resulting in modified motifs. We have studied the behavior of a three-element network module, showing that dynamical relaying leads to zero-lag synchronization even in the presence of coupling delays. This behavior corresponds to a stable isochronous synchronization solution of the dynamics, and is possible irrespective of the distance between the two outer elements, provided the two branches have similar lengths. Our results show that generic dynamical networks can profit from collective synchronization phenomena which can even overcome the limitation of inter-element propagation delays.

Research supported by the Ministerio de Educaci\'on y Ciencia (Spain) and FEDER (projects CONOCE2, AUCOFLUC and LASEA) and by the Generalitat de Catalunya. In addition, I.F. acknowledges funding from the FWO under contract number GP06704-FWOSL21. We thank Wolf Singer for helpful comments and suggestions. We are also thankful to Maxi San Miguel and Dante Chialvo for a careful reading of the manuscript.


\end{document}